# Pinpoint pick-up and bubble-free assembly of 2D materials using PDMS/PMMA polymers with lens shapes

Satoshi Toyoda[1], Teerayut Uwanno[1], Takashi Taniguchi[2], Kenji Watanabe[2] and Kosuke Nagashio[*1]
[1]*Department of Materials Engineering, The University of Tokyo, Tokyo 113-8656, Japan*
[2]*National Institute of Materials Science, Ibaraki 305-0044, Japan*
[*]nagashio@material.t.u-tokyo.ac.jp

**Abstract:** The key to achieving high-quality van der Waals heterostructure devices made by stacking two-dimensional (2D) layered materials lies in having a clean interface without interfacial bubbles and wrinkles. In this study, the pinpoint pick-up and transfer system of 2D crystals is constructed using polymers with lens shapes. We report the bubble-free and clean-interface assembly of 2D crystals in which unidirectional sweep of the transfer interface precisely controlled with the help of the inclined substrate pushes the bubbles away from the interface.

Van der Waals (vdW) heterostructure devices composed of two-dimensional (2D) layered crystals provide a new fundamental platform for many device applications, such as high-performance transistors,[1,2)] light-emitting diodes[3)] and solar cells,[4)] because stacking various layers is possible without considering the lattice mismatch due to the dangling-bond-free layered structure, unlike the case for conventional 3D semiconductors. In the vdW research field, the rotation angle has been recognized as a new parameter for material design since unconventional superconductivity was realized.[5)] vdW heterostructures are generally fabricated from small exfoliated flakes by a manually controlled transfer process using polymers[6-13)] because heteroepitaxial growth of vdW heterostructures by metal organic chemical vapor deposition[14)] or molecular beam epitaxy[15-17)] is still under investigation. The fabrication techniques of vdW heterostructure by the transfer process have been improved by, for example, (i) deterministic transfer by viscoelastic stamping,[9)] (ii) dry transfer using the difference in the thermal expansion between polymers and inorganic materials,[10)] and (iii) pinpoint pick-up for precise position control.[12)]

Recently, a robotic system that automatically searches for exfoliated 2D crystals and assembles them into superlattices inside a glovebox was successfully constructed to overcome the practically impossible repetitive manual stacking of 29 layers.[18)] However, interfacial contaminants, such as hydrocarbon impurities, air, water and so on, are often incorporated as bubbles at the 2D/2D interface.[19-22)] Indeed, this kind of vdW heterostructures including interfacial bubbles are often utilized positively as cells for liquid in nanometer scale for transmission electron microscopy.[23,24)] However, as far as we consider the electron devices, the intrinsic transport properties of vdW heterostructures are expected to be blinded because bubbles work as trap sites or scattering centers. Therefore, bubble-free stacking is strongly required.[25)] Recently, a large bubble-free area has been reported due to the aggregation of bubbles when the layers are brought together in a conformal manner at 110°C during a hot pick-up.[11)] The key is the higher diffusivity of the interfacial contaminants at 110°C than that at room temperature (RT). However, this conformal hot pick-up often resulted in the incorporation of many bubbles, especially for the case in which the two sets of thick 2D crystals are brought together. In this study, pinpoint pick-up and bubble-free transfer of 2D crystals is achieved by combining precise position control of a lens-shaped polymer and unidirectional sweep of the transfer interface using an inclined substrate.

**Figure 1(a)** shows a micromanipulator alignment system for 2D crystals. The polydimethylsiloxane (PDMS, KE106 & CAT-RG, Shin-Etsu Chemical Co.) lens was prepared by dropping a single liquid PDMS droplet on a PDMS solid base/glass slide using a bamboo skewer and by drying it in an inverted position for 24 hours at RT. Inverted drying is preferable because a larger radius of curvature of the PDMS lens can be obtained compared with the upright position, as shown in **Fig. 1(b)**. Moreover, another small PDMS lens formed on the large one, as shown in **Fig. 1(c)**, greatly reducing the contact area to the substrate by retaining the total height of the lens sufficiently. This double lens structure dramatically increases the controllability of pinpoint pick-up by preventing contact with unwanted 2D crystals near the target 2D crystal during the pinpoint pick-up process. Then, polymethylmethacrylate (PMMA, A11, MicroChem

Co.) was spun coated on the PDMS lenses and they were baked at 75°C for 45 min on a hot plate.

For the pinpoint pick-up by the PMMA/PDMS lens, *h*-BN flakes were prepared on a $SiO_2$(90 nm)/$n^+$-Si wafer by mechanical exfoliation. PMMA/PDMS lenses and target *h*-BN flakes on $SiO_2$/Si wafers were assembled in the micromanipulator alignment system under laboratory air conditions and aligned under an optical microscope. Then, the substrate was heated by a Peltier module from the bottom. The heating and cooling by the Peltier module can be easily altered without removing the sample by changing the voltage direction from the DC power supply. The adhesion of PMMA improves at a substrate temperature above the glass transition temperature of PMMA ($T_g$ = ~ 50°C) because the viscosity is substantially decreased.[10)] Each PMMA/PDMS lens and target *h*-BN flake were mechanically brought into contact by the stepping motor with a resolution of 0.125 μm/pulse at the predetermined substrate temperature. Finally, the pick-up was carried out using the thermal shrinkage of PMMA by reducing the substrate temperature to RT. A typical example of pinpoint pick-up of the target *h*-BN flake when the substrate temperature was reduced from 110°C is shown in **Supplementary movie-1**.

**Figure 2(a)** shows the success ratio of the pinpoint pick-up of the *h*-BN flake as a function of the substrate temperature. The pick-up of the *h*-BN flake was quite difficult at temperatures below 70°C, whereas the success ratio was almost 100 % at temperatures above 110°C. However, the success ratio for crack-free pick-up of the *h*-BN flake became very low at high temperatures, as shown in **Fig. 2(a)** and **(b)**. Although 2D layered flakes showed excellent flexible properties below 10 layers, kinks were formed for 20 or more layers during bending.[26)] Since a thickness of ~20 nm is required for *h*-BN flakes to achieve atomically flat surfaces,[1)] the crack formation is a critical issue for pinpoint pick-up. As the substrate temperature increased beyond the $T_g$, PMMA became softer and the adhesion between PMMA and $SiO_2$ increased. When the PMMA detached from the substrate, it was substantially deformed, resulting in severe cracking in the *h*-BN, as schematically shown in **Fig. 2(c)**. This situation was the same for the pick-up by moving down the stage at a constant substrate temperature. The key point to overcoming this issue is that PMMA becomes hard at temperatures lower than $T_g$. Therefore, after the PMMA/PDMS lenses is attached to the target *h*-BN and the $SiO_2$ substrate at a temperature above $T_g$ and they are kept for a while, the substrate temperature is reduced to 55°C below the $T_g$ while the contact is maintained. Then, *h*-BN can be picked up by moving down the stage because PMMA is too hard to deform, as schemat ically shown in **Fig. 2(c)**. By this procedure, the success ratio of the *h*-BN flake target pick-up without any cracks is close to 90 %, as shown by the black solid square in **Fig. 2(a).** Since 2D crystals are generally hydrophobic due to an inert surface without any dangling bonds, the pick-up of 2D crystals by a 2D crystal is much easier than that by a PMMA/PDMS lens. **Supplementary movie-2** demonstrates the successful pick-up of many 2D crystals by initially picking up *h*-BN flakes by adjusting the substrate temperature. The point here is that there is a benefit to the limited contact area of the PMMA lens to the $SiO_2$ substrate for the second pick-up by *h*-BN, which can be achieved by the double lens structure.

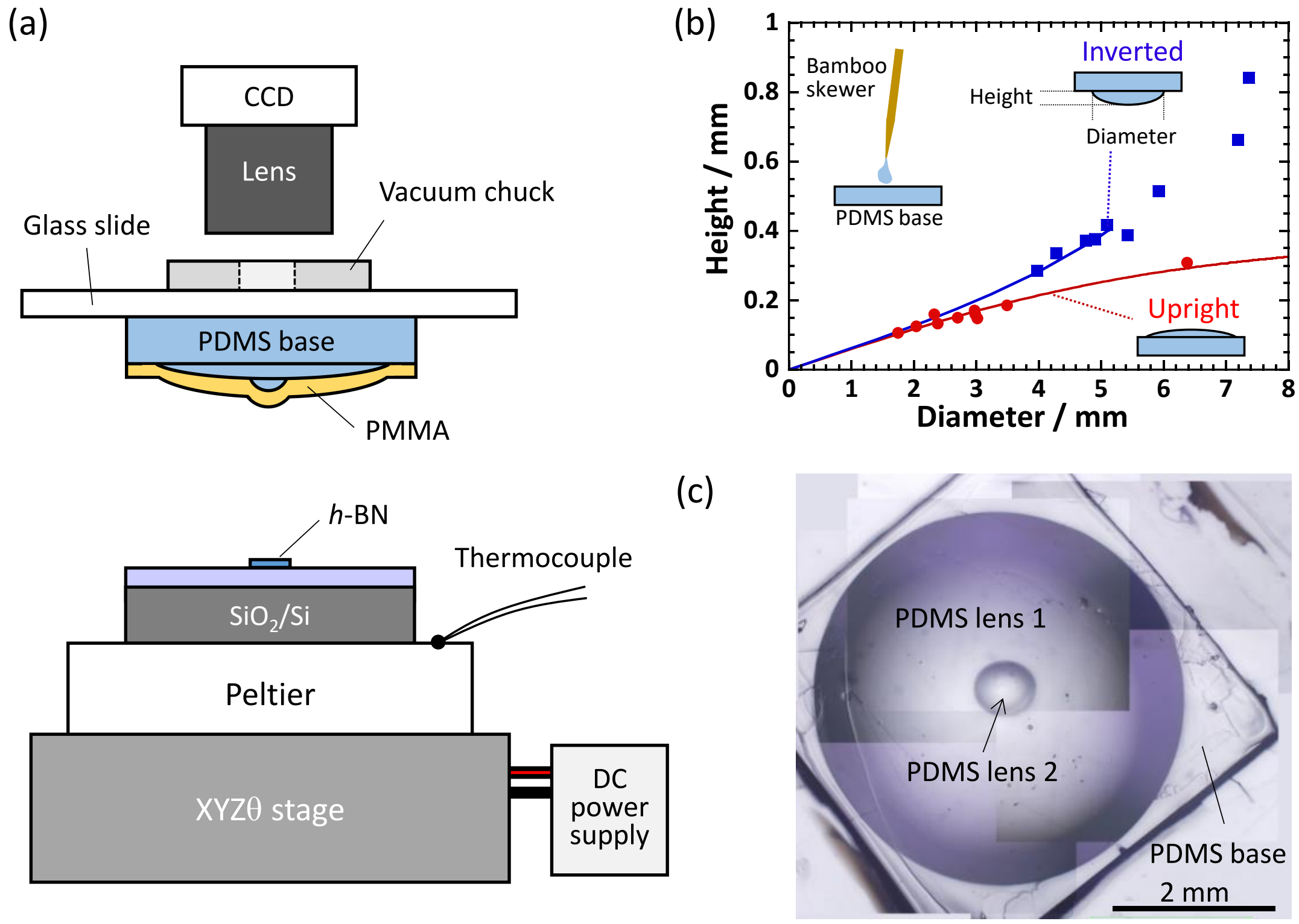

**Figure 1**. (a) Schematic illustration of the micromanipulator alignment system. (b) Relationship between height and diameter for PDMS lens 1 after drying under the inverted and upright states. These data were obtained from the cross-sectional image of the PDMS lens. (c) Photograph of a typical double PDMS lens structure.

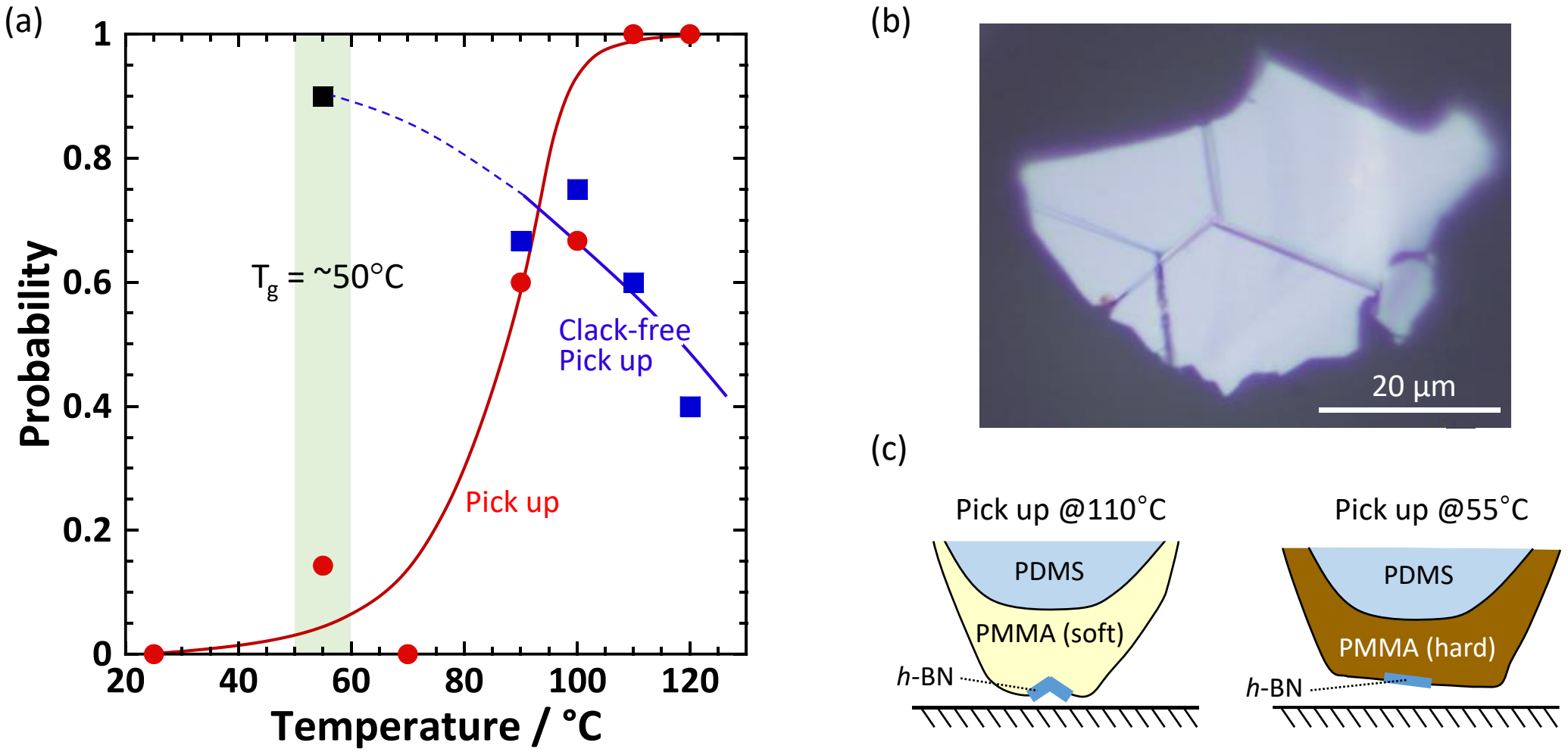

**Figure 2**. (a) Success ratio of pinpoint pick-up of *h*-BN flake (red) as a function of the substrate temperature. The blue solid squares indicate the success ratio of crack-free pick-up per total pick-up. (b) Typical *h*-BN flake (roughly ~50 nm) with cracks picked up at 110°C (>$T_g$). (c) Schematic illustrations for pick-up at 110°C and 55°C. PMMA hardens below $T_g$, which enabled us to pick-up the target *h*-BN flakes without any crack formation.

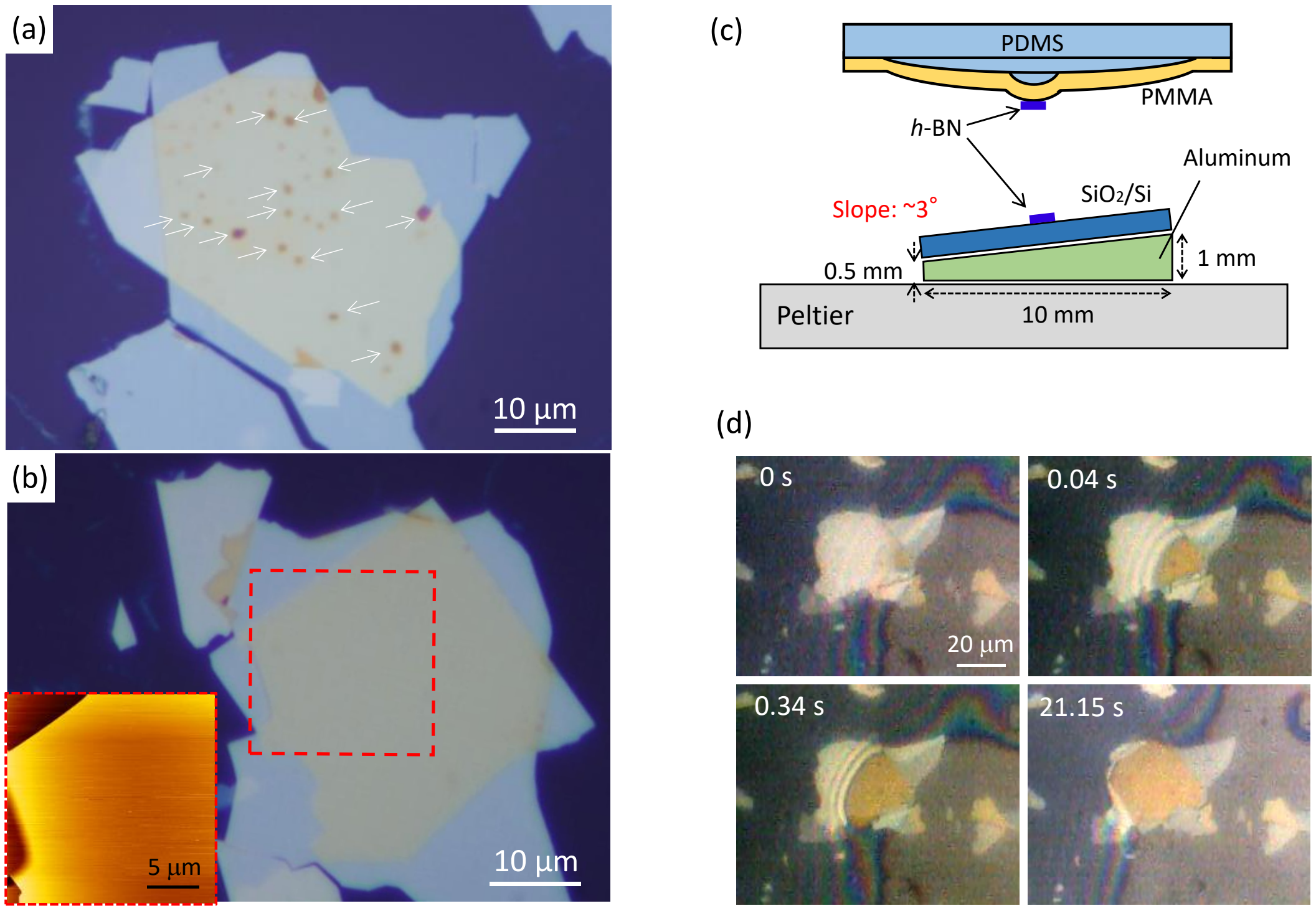

**Figure 3**. (a) Typical image of the *h*-BN/*h*-BN heterostructure obtained by pinpoint pick-up and transfer technique. Many bubbles are observed, as shown by arrows. (b) Typical image of the bubble-free transfer of the *h*-BN/*h*-BN heterostructure. The thicknesses of the top and bottom *h*-BN flakes are 38 nm and 35 nm, respectively, as determined by AFM. The inset shows the AFM image at the red broken rectangular region. (c) Schematic illustration of the inclined substrate and PMMA/PDMS lens. (d) Successive images during the unidirectional transfer using the inclined substrate. This sample is shown in (b).

Now consider the bubble formation at the 2D/2D interface. **Figure 3(a)** shows the typical *h*-BN/*h*-BN heterostructure assembled by the present pinpoint pick-up method. Many bubbles were incorporated at the interface, as shown by arrows. To achieve a bubble-free interface, two kinds of processes can be categorized: one in which the bubbles incorporated are removed from the interface "after the transfer" and the other in which the bubble-free interface is obtained "during the transfer". In our previous study involving the isothermal annealing of a graphene/*h*-BN heterostructure at 200°C after the trasfer,[10] the smaller bubbles aggregated into larger bubbles to reduce the surface energy (Ostwald ripening), which increased the clean interface area. Although the number of bubbles was reduced, the total volume of bubbles was almost the same even after aggregation. Therefore, a completely bubble-free interface was not achieved. Moreover, since the migration of bubbles incorporated at the interface of the two sets of thick 2D crystals is quite limited, obtaining a wide and clean interface area is rather difficult. Therefore, as in the former case, a thermal gradient was produced by local laser heating to enhance bubble migration. As shown in **Supplementary Fig. S1**, two bubbles heated by the Ar laser (red arrows) became large, whereas almost no change was observed for the local heating at the bubble-free region (blue arrows). It was hard to control the migration of bubbles by local laser heating. Another trial involved cyclic annealing, which is known to be effective in removing dislocations from the crystals.[27] As shown in **Supplementary Fig. S2**, a bubble-free interface was not obtained, even though the small bubbles aggregated into larger bubbles. Based on the above trials, the removal of bubbles after the transfer was concluded to be difficult.

Finally, the removal of bubbles during the transfer was considered. The key aspect here was the utilization of the inclined substrate, as shown in **Fig. 3(c)**. Aluminum was used because of its high thermal conductivity. When the *h*-BN flake on the PMMA/PDMS lens contacted the *h*-BN crystal on the $SiO_2$ substrate at 120°C, the contact interface swept from one side to another side, as shown in **Fig. 3(d)** and **Supplementary movie-3**. This unidirectional sweep of the transfer interface pushed the hydrocarbon contaminants, resulting in a bubble-free interface, as shown in **Fig. 3(b)**. The inset image measured by atomic force microscopy (AFM) indicates a bubble-free interface. The double lens structure was important again because the relatively large angle (~3-5°), which cannot be used for planar PDMS stamping, enabled unidirectional sweeping.

In conclusion, pinpoint and bubble-free transfer is achieved by precisely controlling both the hardness of PMMA through the substrate temperature and the sweep direction of the transfer interface using the inclined substrate. This method can be applied to all kinds of 2D crystals.

**Acknowledgements:** We are grateful to Covalent Materials for kindly providing us with Kish graphite. This research was supported by The Canon Foundation, the JSPS Core-to-Core Program, A. Advanced Research Networks, A3 Foresight Program, and JSPS KAKENHI Grant Numbers JP16H04343, Japan.

**References**

1) C. R. Dean, A. F. Young, I. Meric, C. Lee, L. Wang, S. Sorgenfrei, K. Watanabe, T. Taniguchi, P. Kim, K. L. Shepard, and J. Hone, Nat. Nanotechnol. **5**, 722 (2010).
2) T. Uwanno, T. Taniguchi, K. Watanabe, and K. Nagashio, ACS Appl. Mater. Interfaces, **10**, 28780 (2018).
3) F. Withers, O. Del Pozo-Zamudio, A. Mishchenko, A. P. Rooney, A. Gholinia, K. Watanabe, T. Taniguchi, S. J. Haigh, A. K. Geim, A. I. Taratakovskii, and K. S. Novoselov, Nat. Mater. **14**, 301 (2015).
4) M. M. Furchi, A. Pospischil, F. Libisch, J. Burgdörfer, and T. Mueller, Nano Lett. **14**, 4785 (2014).
5) Y. Cao, V. Fatemi, K. Watanabe, T. Taniguchi, E. Kaxiras, and P. Jarillo-Herrero, Nature **556**, 43 (2018).
6) L. Wang, I. Meric, P. Y. Huang, Q. Gao, Y. Gao, H. Tran, T. Taniguchi, K. Watanabe, L. M. Campos, D. A. Muller, J. Guo, P. Kim, J. Hone, K. L. Shepard, and C. R. Dean, Science **342**, 614 (2013).
7) P. J. Zomer, M. H. D. Guimaraes, J. C. Brant, N. Tombros, and B. J. van Wees, Appl. Phys. Lett. **105**, 013101 (2014).
8) R. Yang, X. Zheng, Z. Wang, C. J. Miller, and P. X. -L. Feng, J. Vac. Sci. Technol. B, **32**, 061203 (2014).
9) A. Castellanos-Gomez, M. Buscema, R. Molenaar, V. Singh, L. Janssen, H. S. J. van der Zant, and G. A. Steele, 2D Mater. **1**, 011002 (2014).
10) T. Uwanno, Y. Hattori, T. Taniguchi, K. Watanabe, and K. Nagashio, 2D Mater. **2**, 041002 (2015).
11) F. Pizzocchero, L. Gammelgaard, B. S. Jessen, J. M. Caridad, L. Wang, J. Hone, P. Boggild, and T. J. Booth, Nature Commun. **7**, 11894 (2016).
12) K. Kim, M. Yankowitz, B. Fallahazad, S. Kang, H. C. P. Movva, S. Huang, S. Larentis, C. M. Corbet, T. Taniguchi,

K. Watanabe, S. K. Banerjee, B. J. LeRoy, and E. Tutuc, Nano Lett. **16**, 1989 (2016).
13) L. Banszerus, M. Schmitz, S. Engels, M. Goldsche, K. Watanabe, T. Taniguchi, B. Beschoten, and C. Stampfer, Nano Lett. **16**, 1387 (2016).
14) K. Kang, S. Xie, L. Huang, Y. Han, P. Y. Huang, K. F. Mak, C. -J. Kim, D. Muller, and J. Park, Nature **520**, 656 (2015).
15) M. M. Ugeda, A. J. Bradley, Y. Zhang, S. Onishi, Y. Chen, W. Ruan, C. Ojeda-Aristizabal, H. Ryu, M. T. Edmonds, H. -Z. Tsai, A. Riss, S. -K. Mo, D. Lee, A. Zettl, Z. Hussain, Z. -X. Shen, and M. F. Crommie, Nature Phys. **12**, 92 (2016).
16) T. Hotta, T. Tokuda, S. Zhao, K. Watanabe, T. Taniguchi, H. Shinohara, and R. Kitaura, Appl. Phys. Lett. **109**, 133101 (2016).
17) M. Nakano, Y. Wang, Y. Kashiwabara, H. Matsuoka, and Y. Iwasa, Nano lett, 17, 5595 (2017).
18) S. Masubuchi, M. Morimoto, S. Morikawa, M. Onodera, Y. Asakawa, K. Watanabe, T. Taniguchi, and T. Machida, Nature Commun. **9**, 1413 (2018).
19) S. J. Haigh, A. Gholinia, R. Jalil, S. Romani, L. Britnell, D. C. Ellias, K. S. Novoselov, L. A. Ponomarenko, A. K. Geim, and R. Gorbachev, Nature Mater. **11**, 764 (2012).
20) W. Pan, J. Xiao, J. Zhu, C. Yu, G. Zhang, Z. Ni, K. Watanabe, T. Taniguchi, Y. Shi, and X. Wang, Sci. Rep. **2**, 893 (2012).
21) A. V. Kretinin, Y. Cao, J. S. Tu, G. L. Yu, R. Jalil, K. S. Novoselov, S. J. Haigh, A. Gholinia, A. Mischenko, M. Lozada, T. Georgiou, C. R. Woods, F. Withers, P. Blake, G. Eda, A. Wirsig, C. Hucho, K. Watanabe, T. Taniguchi, A. K. Geim, and R. V. Gorbachev, Nano Lett. 14, 3270 (2014).
22) P. Bampoulis, V. J. Teernstra, D. Lohse, H. J. W. Zandvliet, and B. Poelsema, J. Phys. Chem. C **120**, 27079 (2016).
23) J. M. Yuk, K. Kim, B. Aleman, W. Regan, H. H. Ryu, J. Park, P. Ercius, H. M. Lee, A. P. Alivisatos, M. F. Crommie, J. Y. Kee, and A. Zettl, Nano Lett. **11**, 3290 (2011).
24) Y. Sasaki, R. Kitaura, J. M. Yuk, A. Zettl, and H. Shinohara, Chem. Phys. Lett. **650**, 107 (2016).
25) R. Frisenda, E. Navarro-Moratalla, P. Gant, D. P. De Lala, P. Jarillo-Herrero, R. V. Gorbachev, and A. Castellanos-Gomez, Chem. Soc. Rev. **47**, 53 (2018).
26) D. -M. Tang, D. G. Kvashnin, S. Najmaei, Y. Bando, K. Kimoto, P. Koskinen, P. M. Ajayan, B. I. Yakobson, P. B. Sorokin, J. Lou, and D. Golberg, Nature Commun. **5**, 3631 (2014).
27) Y. H. Tan, and C. S. Tan, Thin Solid Films **520**, 2711 (2012).

**Supplementary figures**

# Pinpoint pick-up and bubble-free assembly of 2D materials using PDMS/PMMA polymers with lens shapes

Satoshi Toyoda[1], Teerayut Uwanno[1], Takashi Taniguchi[2], Kenji Watanabe[2] and Kosuke Nagashio[*1]

[1]*Department of Materials Engineering, The University of Tokyo, Tokyo 113-8656, Japan*

[2]*National Institute of Materials Science, Ibaraki 305-0044, Japan*

[*]nagashio@material.t.u-tokyo.ac.jp

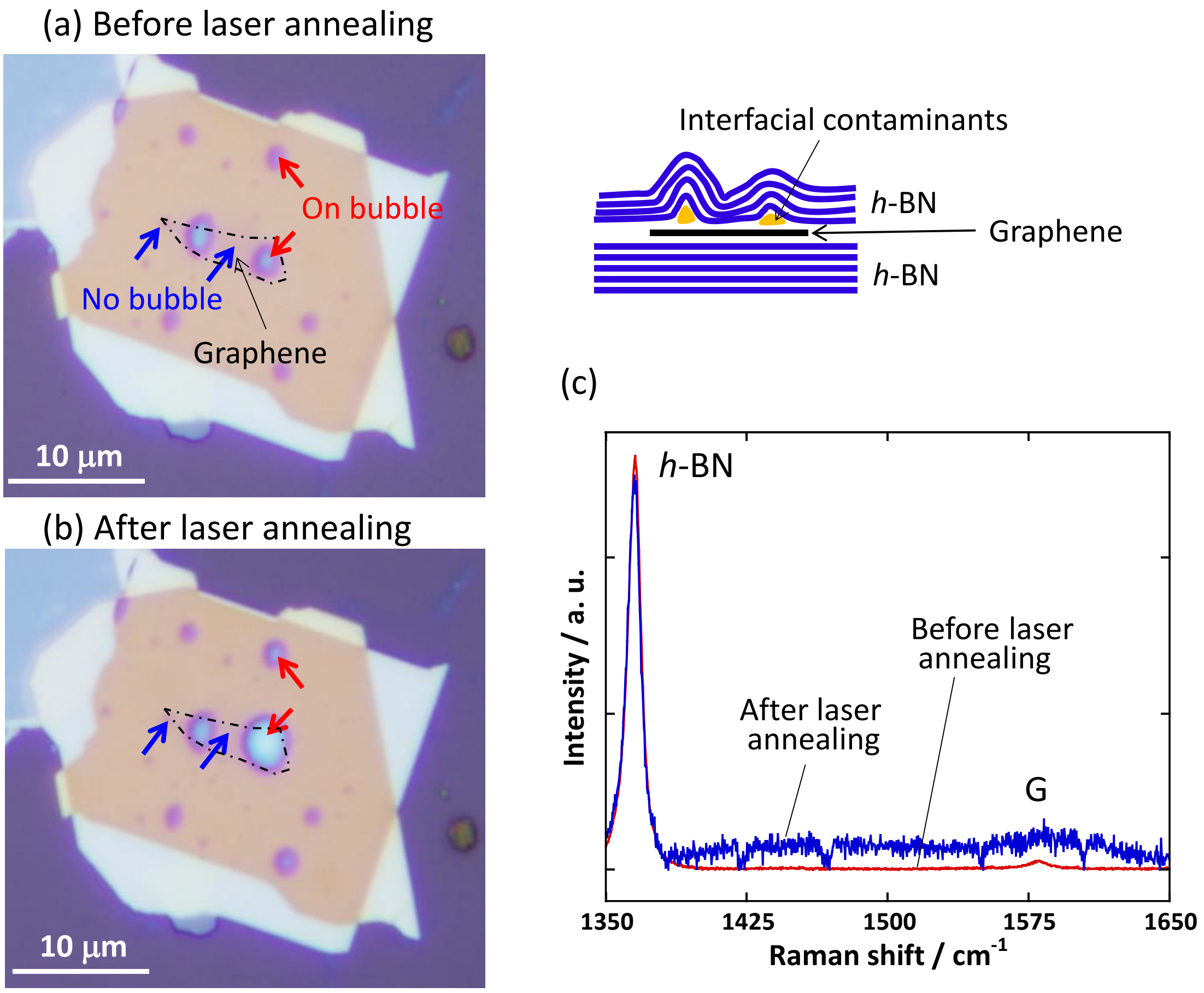

**Fig. S1.** (a) Photograph of the *h*-BN/monolayer graphene/*h*-BN heterostructure with interfacial bubbles before laser annealing. The cross-sectional view is schematically illustrated. Bubbles exist mainly at the *h*-BN/graphene interface. (b) Photograph after laser annealing. The laser annealing was 180 s at 16 mW. The substrate temperature was 120°C. Under this enhanced condition, the aggregation of bubbles was observed, not under low power conditions. (c) Raman spectra of graphene became disordered after 16 mW laser annealing.

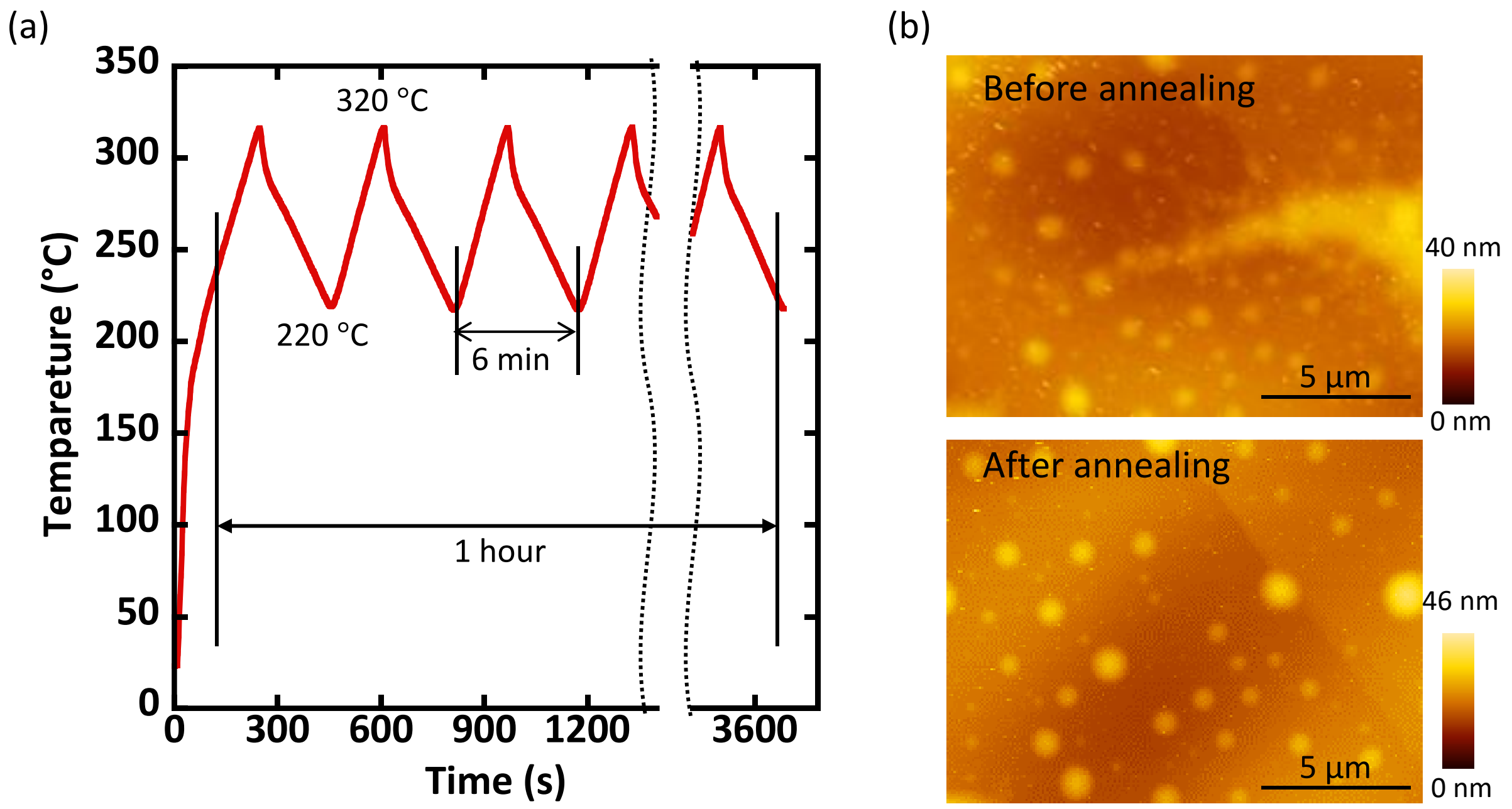

**Fig. S2.** (a) Furnace temperature as a function of time. A rapid thermal annealing system was used for this cyclic annealing. (b) AFM images of the *h*-BN/*h*-BN heterostructure before (top) and after (bottom) cyclic annealing. Although the aggregation of bubbles was clearly observed, a bubble-free interface was not observed.

Information on supplementary movies

**Supplementary movie-1**. A typical example of pinpoint pick-up of the target *h*-BN flake when the substrate temperature was reduced from 110°C. The image of *h*-BN flake after the pick-up is shown in **Fig. 2(b)**. "X8" and "X4" that appeared in the top left corner of the movie indicate eight speed and quad speed, respectively. Image size is 240×210 pixels.

**Supplementary movie-2.** A typical example of successful pick-up of many 2D crystals by initially picking up *h*-BN flakes by adjusting the substrate temperature. Image size is 690×580 pixels.

**Supplementary movie-3**. A typical example of the bubble-free transfer where the unidirectional sweep of the transfer interface pushes the hydrocarbon contaminants. Optical image of resultant bubble-free heterostructures are shown in **Fig. 3(b)**. Image size is 640×480 pixels.